# Visualizing Sound Directivity via Smartphone Sensors

Scott H. Hawley, Department of Chemistry & Physics, Belmont University, Nashville TN
Robert E. McClain Jr., OmegaLab Studio, Nashville TN

## I. INTRODUCTION

When Yang-Hann Kim received the Rossing Prize in Acoustics Education at the 2015 meeting of the Acoustical Society of America, he stressed the the importance of offering visual depictions of sound fields when teaching acoustics[1]. Often visualization methods require specialized equipment such as microphone arrays[2] or scanning apparati[3]. We present a simple method for visualizing angular dependence in sound fields, made possible via the confluence of sensors available via a new smartphone app[4] which the authors have developed.

The versatility of smartphones for use in introductory physics experiments has been demonstrated in several education circles such as in The Physics Teacher and elsewhere[5,6,7,8]. The motivation for the new app described here arose while trying to 'assemble' a set of experiments suitable for an Electroacoustics course to general education students.

## II. EXPERIMENT DESIGN

*2.1. Acoustical Environment*
This app serves as a low-cost stand-in for expensive automated data acquisition systems, as well as a convenient alternative to lengthy manual point-by-point measurements, yet it does not alter the acoustical environment of the experiment space. Industry-standard measurements of loudspeaker and microphone directivity are conducted in anechoic isolation chambers, using 'robotic' systems which coordinate the automated rotation of the product being characterized with the generation and recording of sounds. Such environments and apparati are easily beyond the reach of high schools and small universities. It is however possible to obtain 'serviceable' results, provided a few 'common sense' electro-acoustical criteria are met:

- **Low ambient noise relative to signal**
- **Minimal room reflections**
- **If there are reflections, then stationarity with respect to room modes:** Better results will be obtained if the source and/or recording device are rotated while remaining in the same position, as opposed to 'sweeping out' an arc in space.
- **Minimal reflections off of human operator(s)**

One way to create 'free field' conditions is to conduct the experiments outdoors -- offering students the chance to walk around outside may provide a welcome break from the laboratory. (A wireless Bluetooth speaker could be used as a sound source if running power cables is an issue.) For the results presented here, we used an effectively outdoor environment: a large US Army Lightweight Maintenance Enclosure tent in the forest. Results obtained in a typical laboratory/classroom at a local university are comparable, shown below. To help reject ambient noise and possibly allow multiple simultaneous experiments, we may add a tunable narrow-band filter to the app in the future, but this has not been implemented yet.

*2.2. App Design*
The app is designed to produce a real-time polar plot in which the radius is the sound input level in deciBels obtained either the phone's internal microphone or an external source (via the analog input jack), the angle represents the phone's orientation obtained from its various orientation sensors. It is intended for use with steady-state, single-frequency sound sources. This graph can be exported either via the phone's screenshot capability, or via a .CSV text file written in the app's filesystem which can be transferred to a computer using iTunes File Sharing.

Specifics regarding the app's design and calibration are provided in Appendices A and B.

**III. SAMPLE EXPERIMENTS AND RESULTS**

It is trivial to verify that the iPhone's internal microphone is omnidirectional: generating a steady tone (e.g. by whistling) and spinning the phone while running the app produces what is essentially a perfect circle. It is to be expected that such a microphone would be omnidirectional, i.e. to allow for effective speakerphone usage. For other measurements such as the following, one can use a 3.5mm TRRS adaptor to 'feed' signals from other recording apparati into the phone's analog audio input jack.

*3.1. Loudspeaker*
We placed the phone atop a loudspeaker (Mackie HR824mk2) and co-rotated them on a rotating table, at a distance of 1m from a stationary Earthworks (omnidirectional) calibration microphone (connected via a long cable), as shown in Figure 1a. Figure 2 shows the results for two different test tones.

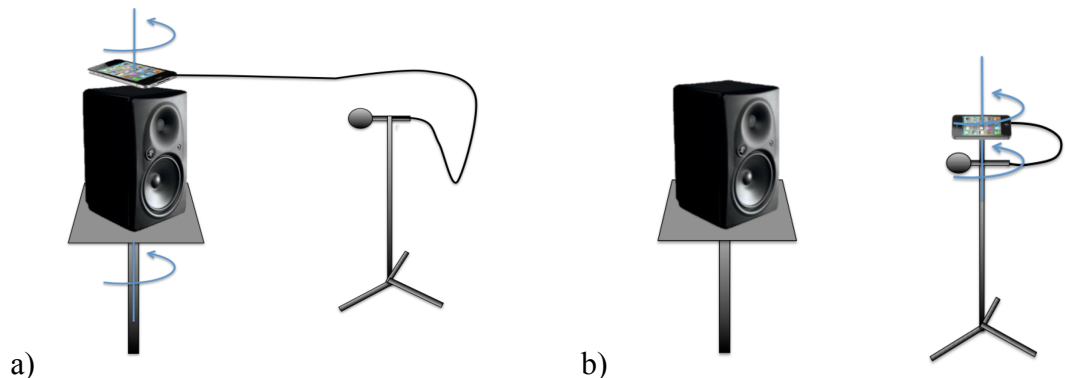

a)  b)
Figure 1: a) Experiment setup for the measurements in of loudspeaker directivity show in in Figure 2. b) Experiment setup for the microphone polar pattern measurement shown in Figure 3; in real life a small, thick towel was rested on the microphone, and the phone rested upon the towel.

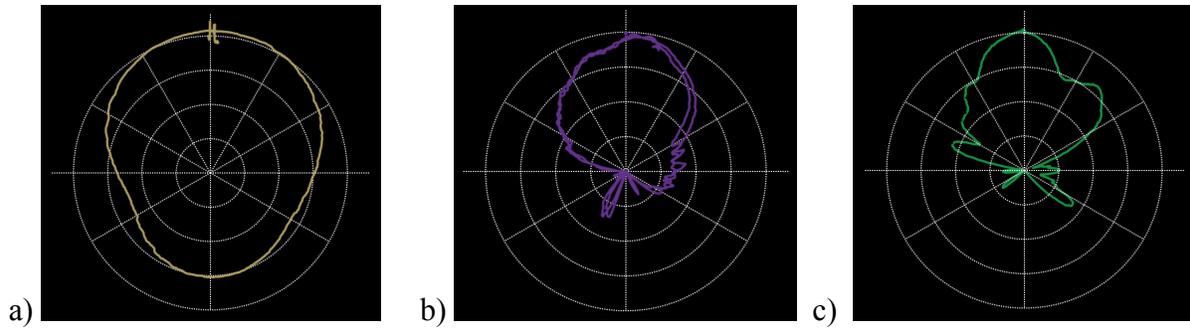

a)        b)        c)

Figure 2: Partial screenshot from "Polar Pattern Plotter" iOS app[4] for measuring directivity of tones at a) 250 Hz and b) 4 kHz generated by a Mackie HR824mk2 studio monitor. c) The same equipment and frequency as b), but this result was obtained in a classroom/laboratory setting in which there are nontrivial room reflections, in contrast to the outdoor conditions of b). Units are dBFS, with the origin being -25dB and the outer ring at 0 dB, and the upward direction corresponding to the front of the speaker.

### 3.2. Cardioid Microphone

A 1kHz tone was played from a speaker at 1 meter from a Shure SM57 dynamic microphone. The (pre-amped) microphone signal was then routed into the audio input jack of an iPhone 5s which was mounted above the microphone along the same axis of rotation. The screenshot shown in Figure 3, resulting from multiple rotations of the microphone shows the classic cardioid pattern.

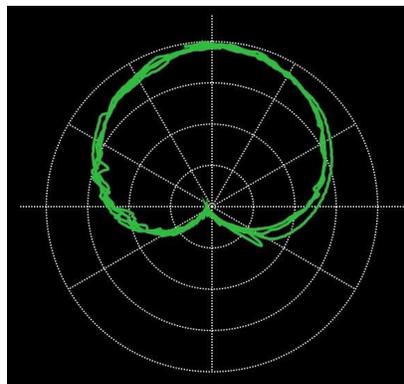

Figure 3: Screenshot of a cardioid pattern obtained for a 1kHZ test tone recorded on a Shure SM57 dynamic microphone. Units are dBFS with the origin at -25dB. Multiple passes are shown; in a future version of the app, an automatic averaging method is to be implemented for producing smoother graphs.

### 3.3. Two-Speaker Interference:

A common acoustics demonstration of wave interference consists of placing two speakers a fixed difference apart, playing a single tone out of both, and allowing students to walk around the room and experience 'loud spots' and 'quiet spots' resulting from the interference. We sought to make this experience visual through the use of our phone app. Results are shown in Figure 4, which provide a visible and quantitative measure illustrating nodal lines and amplitude variations sufficient for acoustics education. We found that holding the phone in the hand, even with a constant-tension cable as a guide for distance and radial direction orientation, offered poor angular accuracy. Instead, suspending the phone via a 'cradle' of tape as shown in Figure 4c was sufficient produce serviceable polar plots.

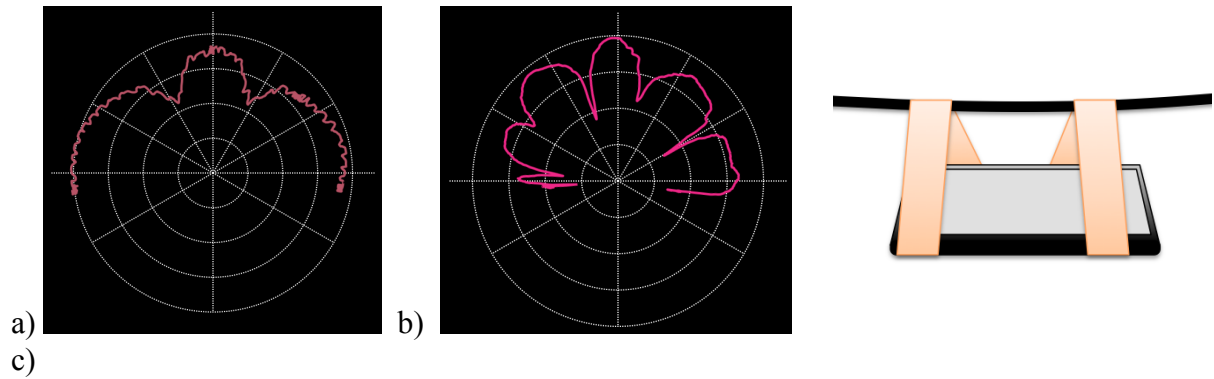

a)
b)
c)

Figure 4: a) & b): Screenshots showing two-speaker interference patterns for pure tones of frequencies (a) 250 Hz and (b) 2500 Hz. Speakers were placed 6 ft apart and measured in an arc of radius 8 ft, by using a microphone cable for constant distance, with the phone suspended via a 'cradle' of tape as shown in (c). The cradle significantly improved angular accuracy compared to simply trying to hold the phone in a radial direction.

Two minor issues arose in the course of these two-speaker interference measurements.

1. Angular 'drift'. W noticed some minor 'drift' of the yaw values provided by the phone when compared to the true physical rotation, usually in the form of under-reporting orientation changes -- e.g. a physical change of 180° registering as only, say 170°. This drift is a known issue with phone-based gyroscope applications[9,10], that the DeviceMotion construct (which we are using) is intended to minimize. The drift is small and does not grow with time, but does grow with increasing angular displacement. This is subject of current investigation; for now, a tunable angular calibration factor has been added to allow the user to compensate for any drift.
2. Sound level fluctuations. In Figure 4a for 250 Hz, we see 'wiggles' in the graph which are not present in Figure 4b) for 2500 Hz. These variations of roughly +/-1 dB appeared each time we repeated the experiment, despite trying to move smoothly and keep the cable at constant tension. It is unclear whether they are an artifact of the measurement process, the software itself, or whether they constitute real (i.e., physical) variations. Recent measurements of large subwoofer arrays by students[11] also show similar fluctuations.

*3.4 Teaching Results*
To date, the use of the app in formal teaching environments has been limited to that of general-education students using early version of this app during (a laboratory exercise for measuring loudspeaker directivity), and a small-group proejct study of subwoofer array patterns[11]. Using the app in the gen-ed context was found to intrinsically and extrinsically support student learning, reported as enthusiasm due to: making it possible to easily visualize the sound; early adoption of new app technology in their class; convenience of real-time automatic-graphing capability; and accessibility of technology (e.g., using personal mobile devices). Students employing the app to study subwoofer array patterns at a large live-sound company's warehouse found the app to offer a convenient alternative to laborious data acquisition, and the graphing capability afforded immediate feedback of experimental results.

IV. CONCLUSION

The combination of sound recording, orientation sensing, and graphical analysis offered by a smartphone app has made it possible to perform measurements of sound directivity (e.g. polar patterns) without the need for specialized equipment, in a way that is suitable for introductory students in acoustics. The app is published and available for free on the iTunes Store[4]. The graphs obtained provide sound visualization, which agree with both qualitative human experience as well as manufacturers' specifications. Student and educator response has so far been limited but positive.

Several areas for future work include investigating the drift reported in Section 3.4, and the use of the accelerometer for measuring sound directivity in vertical planes. We also would like to investigate a narrow-band filter for noise rejection and allowing multiple experiments to occur together, as well as a possible chirp-timing algorithm for rejecting reflections.

Other educators are invited to use the app to develop their own experiments and share them. To facilitate the improvement of this software as a community tool, we have made the app free to download and the source code available[4].

___________________________________________________________________________

The following appendix is intended to be an external online resource.

**APPENDIX A: APP DESIGN SPECIFICS**
Sound level measurements can vary by +/- 10 deciBels (dB) between various smartphone devices and apps[1], so we did not presume to conduct a detailed calibration study of acoustic sound pressure for this app.  Furthermore, since the app is likely to offer best results when used with external microphones and recording systems which may present different signal levels to the phone, any default calibration would be meaningless.  We thus made a rough calibration, where we mapped values obtained from the iPhone 5's microphone to those from an SPL meter.  For further calibration of sound level, the app provides the user with an interactive "slider" which offers +/- 15dB variance.  Details of calibration tests are available in Appendix B.

Modern professional measurements also make use of frequency-swept signals[2].  For simplicity, the current app only measures steady state sources and does not perform any frequency-separating analysis.  Inclusion of such features is a topic for future work.

To obtain the horizontal orientation of the phone, we initially wrote the app to use only the internal compass, but found it to be somewhat unreliable --- even apart from any proximity effects due to the large permanent magnets in loudspeakers.   The use of (time-integrated) gyroscope data for angular orientation has been shown to offer improvements over compass data[3], however uncertainties in gyroscope-based orientation data have also been documented[4].  Thus the app also implements a drift-correction parameter which the user can tune.

For greater accuracy, Apple provides a construct called DeviceMotion[5], which combines compass, gyroscope and accelerometer data.  Our app uses DeviceMotion to provide the rotation angle in the horizontal plane ("yaw") and the vertical plane. In the interest of simplicity, the current version simply uses all three types of readings (compass, DeviceMotion-processed gyroscope yaw,  and DeviceMotion-processed accelerometer) independently, and allows the user the option to display one or more of these data streams when plotting on the screen.

The data from all three sensors is saved as a 4-column .CSV file (as "dB, Angle-Compass, Angle-Gyro, Angle-Accel") in the app's Documents directory on the phone, which can then be transferred to a computer via a USB cable, and could be combined, analyzed or modified in a post-processing manner of the user's choosing, e.g., to enhance the accuracy of angular position or to smooth the data.

**APPENDIX B: SOUND LEVEL CALIBRATION**
The "Polar Pattern Plotter" app provides a rudimentary sound level calibration slider which simply adds a constant to the decibel value provided by the phone's internal system.  Obtaining accurate polar pattern curve shapes does however require that the input sensor(s) are at least linear with respect to sound level.

*B.1 SPL Calibration (iPhone 5s MEMS Microphone)*

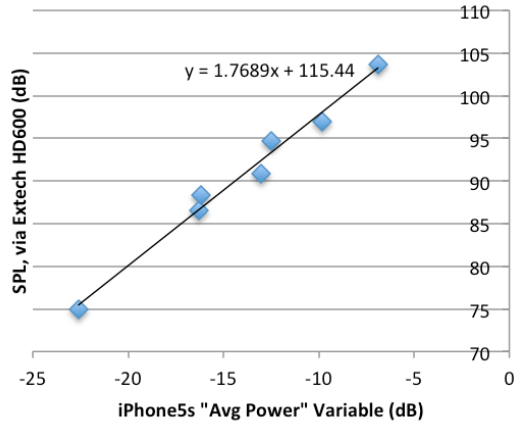

Figure B1: Calibration test for tones at 1 kHz, mapping between the phone's "average power" data provided in software to the physical (external) SPL value as measured on a professional sound level meter (Extech HD600, C weighted, slow).  In the range 75-105 dB the response is essentially linear, although for but below 75dB (not shown) the values from the phone are higher than the linear fit would suggest.

*B.2 Direction Calibration and Comparison to 'Control'*
We made recordings from the microphones into a Digital Audio Workstation, while using a digital motor from a cinema camera mount to obtain uniform rotation rate.  This provided us with a linear mapping between recording time and angular position.  Screenshots of the waveform envelopes were then taken and manually digitized into curves using the software utility PlotDigitizer[6].

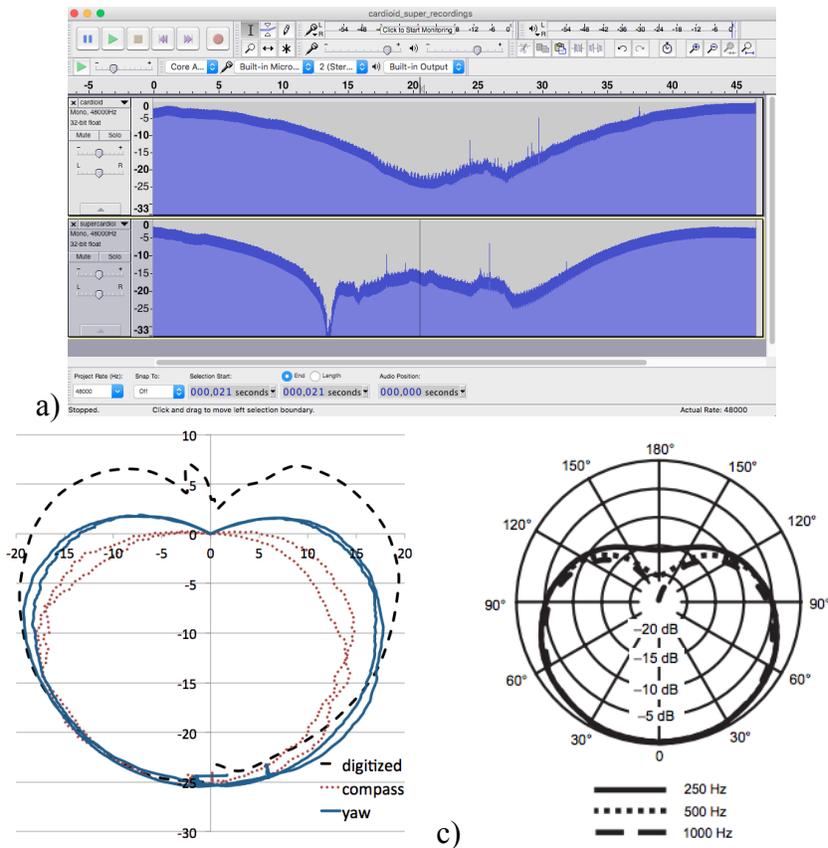

Figure B2: a) Waveform graphs for one complete rotation at constant rate, using cardioid (top) and supercardioid (bottom) settings of a KSM9 condenser microphone, for a 1kHz test tone. Screenshots of the 'envelope' contours of these waveforms were taken, and then digitized into curves and plotted in Excel, resulting in the long-dashed shown in b). Data obtained via the app for the cardioid setting is also shown in b), which shows a comparison between using the compass for directional orientation (dotted line), versus using the "yaw" output from the DeviceMotion construct (solid line). The latter is shown to offer superior results when compared to the manufacturer's specifications[7] in c) (long-dashed line), and was used for all other measurements in this paper.